\documentclass[aps,prl,twocolumn,letterpaper,superscriptaddress,showpacs]{revtex4-1}
\usepackage[colorlinks,linkcolor=blue,anchorcolor=blue, citecolor=blue,urlcolor=blue,]{hyperref}
\usepackage{graphicx}
\usepackage{amssymb}
\usepackage{CJKutf8}
\usepackage{verbatim}
\usepackage{physics}
\usepackage{amsmath}
\usepackage{dirtytalk}
\usepackage{soul}
\providecommand{\bibnamefont}[1]{#1}
\providecommand{\bibfnamefont}[1]{#1}

\begin{document}

%\begin{CJK*}{UTF8}{bsmi}
\begin{CJK*}{UTF8}{gbsn}

\title{Transport in the emergent Bose liquid: \\
Bad metal, strange metal, and weak insulator, all in one system}
\author{Tao Zeng (\CJKfamily{gbsn}曾涛)}
\affiliation{Tsung-Dao Lee Institute \& School of Physics and Astronomy, Shanghai Jiao Tong University, Shanghai 200240, China}

\author{Anthony Hegg}
\affiliation{Tsung-Dao Lee Institute \& School of Physics and Astronomy, Shanghai Jiao Tong University, Shanghai 200240, China}

\author{Long Zou (\CJKfamily{gbsn}邹龙)}
\affiliation{Tsung-Dao Lee Institute \& School of Physics and Astronomy, Shanghai Jiao Tong University, Shanghai 200240, China}

\author{Shengtao Jiang (\CJKfamily{gbsn}蒋晟韬)}
\affiliation{Tsung-Dao Lee Institute \& School of Physics and Astronomy, Shanghai Jiao Tong University, Shanghai 200240, China}

\author{Wei Ku (\CJKfamily{gbsn}顧威)}
\altaffiliation{corresponding email: weiku@mailaps.org}
\affiliation{Tsung-Dao Lee Institute \& School of Physics and Astronomy, Shanghai Jiao Tong University, Shanghai 200240, China}
\affiliation{Key Laboratory of Artificial Structures and Quantum Control (Ministry of Education), Shanghai 200240, China}
\affiliation{Shanghai Branch, Hefei National Laboratory, Shanghai 201315, China}

\date{\today}

\begin{abstract}
Non-saturating high-temperature resistivity (``bad metal''), $T$-linear low-temperature resistivity (``strange metal''), and a crossover to activation-free growth of the resistivity in the low-temperature limit (``weak insulator'') are among the most exotic behaviors widely observed in many strongly correlated materials for decades that defy the standard Fermi liquid description of solids. Here we investigate these puzzling behaviors by computing temperature-dependent optical conductivity of an emergent Bose liquid and find that it intuitively explains them all, including a featureless continuum and a well-known mid-infrared peak. Interestingly, even the observed carrier density dependence of the resistivity has an analytical explanation.
The unification of all these non-Fermi liquid behaviors in a single model suggests that a new quantum state of matter, the emergent Bose liquid, offers another general paradigm alternative to the Fermi liquid.
\end{abstract}

\maketitle

\end{CJK*}

Transport properties are among one of the most important features of materials widely used in many applications.
Typical solid state physics textbooks start with the Drude model that provides a simple description of the transport properties of metals and lays the foundation of the Fermi liquid theory.
It is therefore quite unsettling that, in strongly correlated materials, several atypical characteristics have been repeatedly observed for decades and yet consensus of reasonable understanding is still lacking.
In essence, these exceptional features reflect the important fact that these materials are in some novel quantum states whose response defies the standard lore.

\begin{figure}
	\centering
	\includegraphics[width=\columnwidth]{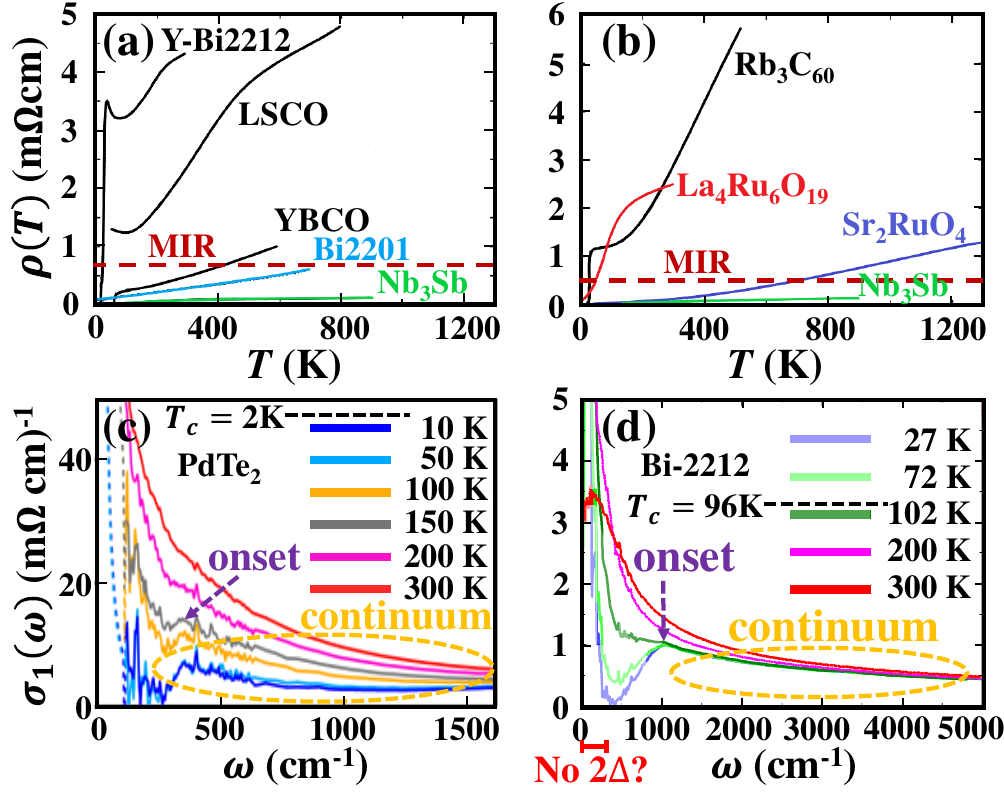}
    \vspace{-0.6cm}
	\caption{Temperature-dependent resistivity for several (a) cuprate~\cite{wang_observation_1996,PhysRevB.54.7445,PhysRevLett.69.2975,PhysRevB.42.6342,PhysRevB.41.846} and (b) other strongly correlated materials~\cite{Hebard1993,Khalifah2001,Tyler1998} hosting characteristic $T$-linear resistivity ranges and complete omission of the MIR saturation bound~\cite{Gunnarsson} (red dotted lines) respected by Fermi liquid such as $\text{Nb}_3\text{Sb}$~\cite{Fisk1976}.
    (c) Optical conductivity of PdTe$_{2}$ with ever-present continuum (orange dotted circle) and a temperature-independent mid-infrared peak visible at low temperature (purple dotted arrow).
    (d) Optical conductivity of Bi-2212~\cite{hwang2007} with strongly similar features to (c) apart from a different overall scale.}
	\label{exp_figs}
    \vspace{-0.5cm}
\end{figure}

One well-known example is the so-called ``bad metal'' behavior~\cite{PhysRevLett.74.3253,Gunnarsson}, namely a non-saturating resistivity at high temperature ($\sim300$K) beyond the Mott-Ioffe-Regel (MIR) limit~\cite{Hussey2004}, found in many strongly correlated materials such as cuprates~\cite{wang_observation_1996,PhysRevB.54.7445,PhysRevLett.69.2975,PhysRevB.42.6342,PhysRevB.41.846,Takenaka}, vanadates~\cite{Qazilbash2006}, ruthenates~\cite{Khalifah2001,Tyler1998}, and
Rb$_3$C$_{60}$~\cite{Hebard1993}, to give a few examples in Fig.~\ref{exp_figs}(a)(b).
A naive textbook interpretation would result in an unphysical mean-free-path shorter than atomic size, defying the large intra-atomic energy spacing.

Another example of atypical transport is the so-called ``strange metal'' behavior~\cite{Legros2018}, namely a temperature $T$-linear resistivity at low temperature, discovered in cuprates~\cite{wang_observation_1996,PhysRevB.54.7445,PhysRevLett.69.2975,PhysRevB.42.6342,PhysRevB.41.846,Takenaka,daou_linear_2009,cooper2009,Legros2018}, twisted bi-layer graphene~\cite{PhysRevLett.124.076801}, iron pnictides~\cite{Kasahara2010}, ruthenium oxides~\cite{grigera_magnetic_2001}, organic metals~\cite{PhysRevB.80.214531}, transition-metal dichalcogenides~\cite{VSe2_resist2020,TaS2_resist2008} and heavy-fermion compounds~\cite{gegenwart_quantum_2008}. [e.g., see $T$-linear resistivity of several materials in Fig.~\ref{exp_figs}(a)].
Such $T$-linear resistivity is disabled by the Pauli principle of fermionic quasiparticles, so it strongly suggests that the core foundation of solid-state physics is completely broken.

A third peculiar transport property at lower temperature is the so-called ``weak insulating'' behavior, where the metallic reduction of the resistivity crosses over to an insulator-like growth in the low-temperature limit.
Most noticeably, this low-temperature resistivity growth $\rho(T)$ cannot be described by $e^{\Delta/T}$ as in typical insulators, indicating the lack of an activation energy scale $\Delta$.
[Some experimental data appears similar to $\ln{(1/T)}$~\cite{PhysRevLett.75.4662,PhysRevLett.85.638}.]
Such a weak insulating behavior has been widely observed in cuprates~\cite{PhysRevLett.75.4662,PhysRevLett.85.638}, twisted bi-layer graphene~\cite{PhysRevLett.124.076801}, nickelates~\cite{li_absence_2020}, iron pnictides~\cite{PhysRevB.79.212510}, transition-metal dichalcogenides~\cite{VSe2_resist2020} and vanadium oxyselenides~\cite{WenHaihu2018}.
It is very puzzling how a coherent gapless fermionic quantum system displays smaller resistivity (weaker scattering) upon increasing temperature.

Perplexingly, the optical conductivity, which provides more complete dynamical information for resolving these puzzles, suffers from yet more mysteries, as shown in Fig.~\ref{exp_figs}(c)(d).
First, despite the clean gaps between bands, there is a mysterious overwhelming continuum exhausting most of the spectral weight.
Second, many of these materials~\cite{hwang2007,Startseva,Basov} show a clear doping- and temperature-independent mid-infrared onset ($\sim 100$meV) of no obvious origin.
Finally, the optical conductivity displays gap-less excitations without even a hint of a superconducting gap at $2\Delta_{SC}$ despite its observation in quasi-particle spectra.

Armed with the knowledge that such unconventional phenomena are incompatible with theories reliant on fermionic quasiparticles, by far the dominant strategy to defeat conventional transport~\cite{DMFT1,PhysRevB.91.075124,PhysRevB.94.235115,PhysRevLett.122.186601,faulkner_strange_2010,RevModPhys.68.13,MottQC_2015PRL,SYKpnas, Subir_PRX,Planckian_RevModPhys, cryst2022, Ataei2022,alexandrov_logarithmic_1997,Europhysics2008} utilizes some form of exotic scattering to disrupt the particle nature.
%while retaining the fermionic description of the theory.
Popular theories for the bad metal primarily incorporate thermal destruction of quasiparticles in strongly interacting systems such as via the Hubbard model ~\cite{PhysRevB.91.075124,RevModPhys.68.13,DMFT1,MottQC_2015PRL,PhysRevB.94.235115}, modified Hubbard model~\cite{PhysRevLett.122.186601}, incoherent metal~\cite{Hartnoll2015} and Mott quantum criticality~\cite{MottQC_2015PRL}.
However, the local interaction scale is much too large ($\sim 8$eV) to generate the low-energy features (e.g. the $\sim100$meV onset of a continuum) in the optical conductivity.
Breaking the quasiparticle nature is also the primary strategy in strange metal theories, which is typically achieved by introducing enough exotic scattering such as in marginal Fermi liquid (MFL)~\cite{varma}, anti-de Sitter space-conformal field theory (AdS-CFT) correspondence~\cite{faulkner_strange_2010,Sachdev_2010}, Sachdev-Ye-Kitaev (SYK) model~\cite{SYKpnas, Subir_PRX}, Planckian dissipation~\cite{Planckian_RevModPhys, cryst2022, Ataei2022} or quantum criticality~\cite{Sachdev1,sachdev_quantum_2011,mousatov_theory_2020,Caprara2022,Sachdev2023,YSYK2024}.
However, observations indicate that $T$-linear resistivity often does not persist down to the lowest temperature scales, casting doubt on the applicability of such theories to a wide class of strongly correlated materials~\cite{Bozovic1}.
Finally, the weak insulating behavior has been studied in the context of disorder scattering via bipolarons~\cite{alexandrov_logarithmic_1997}, Kondo-like physics~\cite{Europhysics2008}, and pseudogap proximity~\cite{laliberte2016origin}.
However, the wide range of temperature and doping where weak insulating behavior is found and its persistence in ever-cleaner samples~\cite{Bozovic1} suggests that it is due to an intrinsic property of these systems over an equally wide parameter range.

Here we propose a different approach in which we retain the particle nature and instead replace the fermionic carriers with bosonic ones that emerge at some high energy scale.
This choice is consistent with recent shot noise experiments on some of these materials~\cite{2eShotBoz,Bastiaans2eShot} that concluded the presence of charge $2e$ carriers in the absence of superconductivity.
Recent Monte Carlo study~\cite{Gaffar2e} also confirms the presence of such emergent bosonic carriers even in the presence of low-energy fermionic carriers.
Specifically, we study the \textit{generic} temperature-dependent dynamical conductivity of an emergent Bose liquid (EBL) theory~\cite{weiku1,weiku2,laurence1,hegg1,Lang2025review}.
As a demonstration, we apply this theory to a square lattice of atoms and find that the result \emph{simultaneously} displays  semi-quantitatively all of the phenomena described above: bad metal, strange metal, weak insulator, and all the essential features of the optical conductivity.

The conditions resulting in an EBL~\cite{weiku1,weiku2,laurence1,hegg1,Lang2025review,xzhang} are quite general.
It begins with a strong local physics that prohibits double occupation of carriers in each atom.
Further assuming \textit{strong} emergent nearest-neighboring correlations that effectively bind fermionic carriers to form bond-center bosons, which conduct pivoting motion in the bond lattice with an extended hardcore constraint to avoid the above double occupation.
Due to the high energy scale of the binding, say higher than 100meV, such bosonic charge-2$e$ carriers are thus robust and unaffected by physics of lower energy.
This is in great contrast to the weakly bound Cooper pairs that are intimately tied to the presence of superconductivity.

An essential feature of such an EBL is the multi-orbital nature associated with the bond orientations in 2D and 3D systems, making EBL sensitive to geometrical frustration.
This enables a novel homogeneous quantum Bose metal phase~\cite{hegg1} with a pseudogap ~\cite{xinlei1} in the fermionic spectral function, and offers a natural explanation to the lack of superconductivity in single-layer graphene\cite{xzhang}, in contrast to its derived superconducting materials.

\begin{figure}
	\setlength{\belowcaptionskip}{+0.0cm}
	\begin{center}
	\includegraphics[width=\columnwidth]{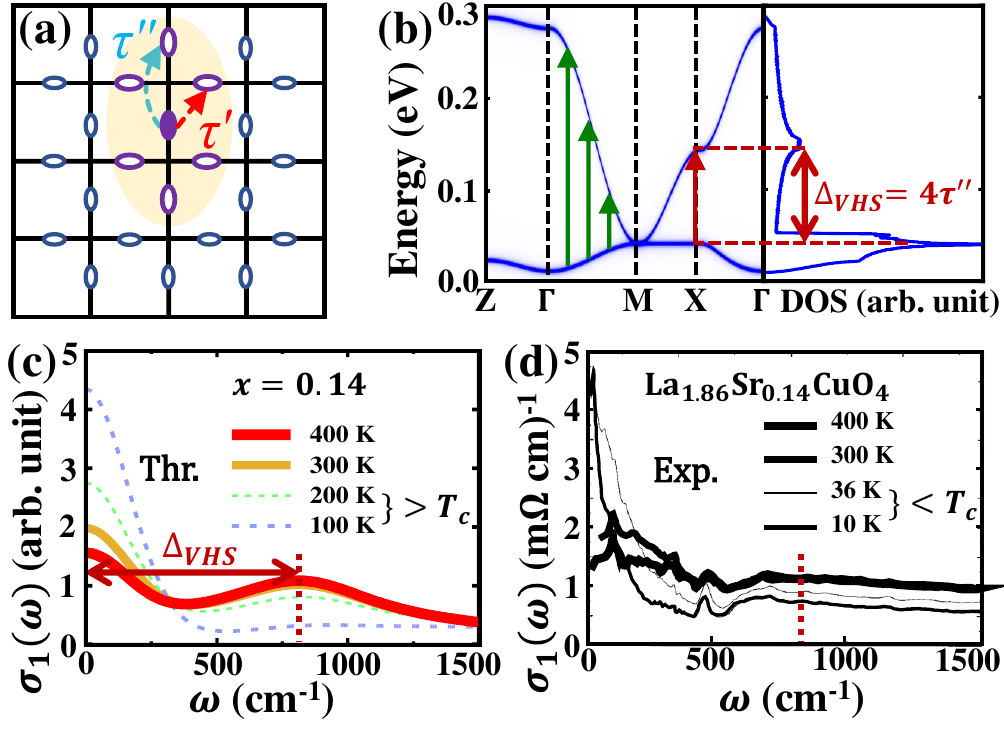}
	\end{center}
    \vspace{-0.5cm}
	\caption{(a) Illustration of the checkerboard bond lattice of emergent bosons as bound neighboring holes that conduct pivoting motion via $\tau^\prime$ and $\tau^{\prime\prime}$.
    Notice the emergent two-orbital nature associated with bonds of different orientations.
    (b) The corresponding band structure and density of states hosting two van Hove singularities at the momentum X=($\pi$, 0) with energy difference $\Delta_{VHS}=4\tau^{\prime\prime}$ (red arrows).
    Notice a continuous supply of gapless inter-band excitations (green arrows).
    (c) The resulting optical conductivity in turn contains a mid-infrared peak at $\Delta_{VHS}$ (red dotted lines) and a gapless continuum.
    (d) Both features are in semi-quantitative agreement with experimental observation in LSCO~\cite{Startseva}, including their temperature dependence.}
	\label{opt_cond_fig}
    \vspace{-0.4cm}
\end{figure}

As an example we consider a simple square lattice of fermions (as found in the cuprates, nickelates, and ruthenates). Following the EBL construction above, the corresponding bond-centered lattice of bosons~\cite{weiku1} formed a checkerboard lattice illustrated in Fig.~\ref{opt_cond_fig}(a)
\begin{eqnarray}
\label{model1}
H &=& \sum_{<l,l'>} \tau_{ll'} b_{l}^\dagger b_{l'}
\end{eqnarray}
where $b_{l}$ denotes the annihilation of a boson located at bond $l$ with corresponding extended hardcore constraint, $b_{l}^\dagger b_{l^\prime}^\dagger=0$ for all first and second neighboring $l$ and $l^\prime$, [illustrated by the yellow region in panel (a)], and $\tau_{ll'} = \tau'$ or $\tau''$ are the dominant kinetic process (without virtually breaking up the boson) between nearest or second nearest neighboring bonds, as previously extracted~\cite{weiku1} from ARPES dispersion data.
In that particular example, $\tau' > \tau'' > 0$ is of the order of $40$meV (for example, $\tau'=33$meV and $\tau''=26$meV~\cite{weiku1}).
Due to partial geometric frustration~\cite{hegg1} between these two hopping paths, the corresponding bandwidths for the lower and higher bands are $\sim 30$meV and 300meV, respectively.
We then evaluate the optical conductivity above the superfluid transition temperature, $T_c$, via the Kubo formula~\cite{Kubo1957}
\begin{eqnarray}
\label{optcond}
\sigma_{\alpha\beta}(\omega) &=& \frac{i}{\omega} [\Pi_{\alpha\beta}(\omega) + \frac{n_0e^2}{m}\delta_{\alpha\beta} ],
\end{eqnarray}
where $\Pi_{\alpha\beta}(\mathbf{q},\tau)=-\frac{1}{\nu}\langle T_{\tau}j_{\alpha}^\dagger (\mathbf{q},\tau) j_{\beta} (\mathbf{q},\tau)\rangle$ is the current correlation function along the $\alpha$ and $\beta$ directions.

Notice that in Fig.~\ref{opt_cond_fig}(d) and other bad/strange metals~\cite{supplementary}, experimental spectra typically display a large and nearly temperature-independent broadening above $T_c$.
This directly forces one to seek \textit{alternative} sources of temperature dependence for transport. 
%To satisfy this stringent observation enforced requirement, 
We thus absorb the influence of the hardcore constraint on dynamical transports into a temperature- (and density-) \textit{independent} effective scattering rate $\eta$ that broadens the spectrum of $\Pi_{\alpha\beta}(\omega)$ computed without interactions.

Such a temperature- and density-independent scattering rate is expected from the bosonic nature of the carriers.
Unlike the Pauli principle-limited phase space of fermionic carriers, the entire phase space is available for the scattering processes of bosonic carriers.
Indeed, the scattering rate of hard-core bosons was found to be \textit{temperature-independent} and of the order of the bandwidth~\cite{badmetHCB}.
Similar large and temperature-independent scattering rate was also found in quantum Monte Carlo simulations~\cite{MGaffar_Jie}.
We thus focus on the essential temperature/density dependent spectral features, by fixing the corresponding $\eta_{1}\sim 30$meV and $\eta_{2}\sim 10\eta_{1}$ for the narrow lower band and the wider higher band, respectively.

Figure \ref{opt_cond_fig}(c) illustrates our result, showing phenomenal favorable agreement with experiments above $T_c$ in panel (d). In particular, we find all of the puzzling phenomena in the optical conductivity discussed above, including a featureless continuum spanning a large frequency range, a mid-infrared feature around $100$meV, and lack of superconductivity-related $2\Delta_{SC}$ feature.

The EBL effortlessly provides an explanation for all of these exotic puzzles.
A broad featureless continuum
%over the entire frequency range
originates from the gapless inter-band transitions [green arrows in Fig.~\ref{opt_cond_fig}(b)].
%It is gapless because the two bands touch due to symmetry and all bosons participate due to a lack of a Pauli principle.
Conspicuously, a mid-infrared feature $\sim 100$meV rises out of this continuum from the transitions between the two van Hove singularities (red arrow), an emergent energy scale of EBL.
Interestingly, this scale depends very weakly to density, in agreement with the observation~\cite{hwang2007}.
Note that both features above are made possible by the naturally emerged multi-orbital characteristic of EBL.
Finally, the lack of a signature for $2\Delta_{SC}$ is self-evident, since bosonic superfluidity does not require a gap in the bosonic band structure.

\begin{figure}
	\setlength{\belowcaptionskip}{+0.0cm}
	\begin{center}
		\includegraphics[width=\columnwidth]{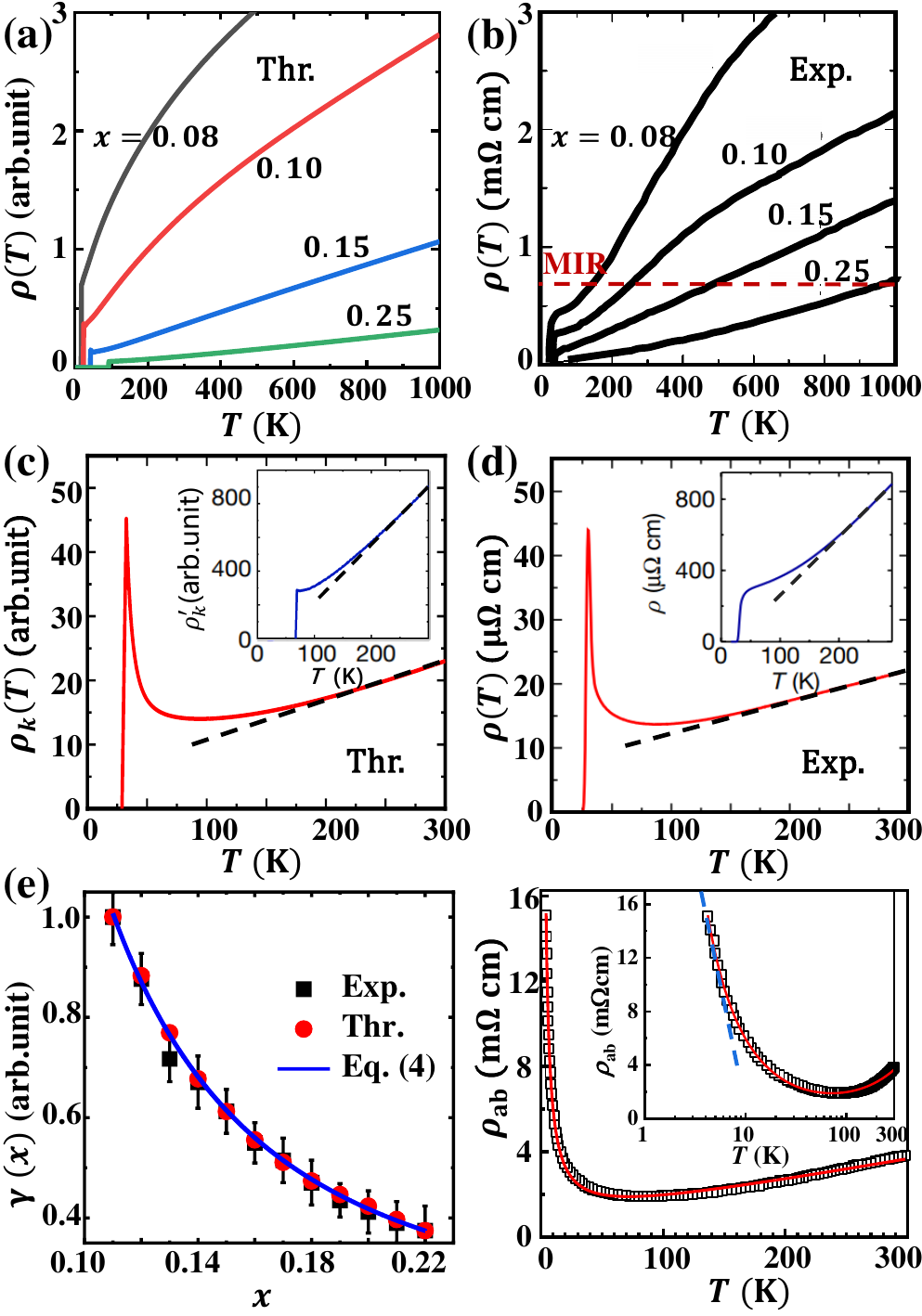}
	\end{center}
    \vspace{-0.5cm}
	\caption{(a) Temperature($T$)- and density($x$)-dependent resistivity of the EBL and (b) the experiments~\cite{Takenaka} showing generally good agreement.
    In particular, both share linear resistivity over a broad range of temperatures, omission of the MIR bound, and general carrier density dependence.
    (c) At lower temperatures, the linear resistivity gives way to weak insulating behavior as observed in (d) experiments~\cite{Bozovic1}.
    (e) Careful analysis of the density dependent resistivity slope agrees with observations~\cite{PhysRevLett.93.267001} up to a density-independent shift accounting for e.g. phonon scattering.
    (f) The Drude weight in the EBL (red line) provides a good phenomenological fit to the experimental in-plane resistivity of LSCO~\cite{Ando_weakIns_2001} over a broad temperature range, superior to the low-temperature $\ln{(1/T)}$ fit (dashed blue line).}
	\label{dc_resist}
    \vspace{-0.5cm}
\end{figure}

Having unveiled all the puzzles in the optical conductivity, we now examine the $\omega \rightarrow 0$ resistivity above $T_c$.
Interestingly, the results in Fig.~\ref{dc_resist}(a) resembles well the experiments in (b) over a wide range of temperature and \textit{carrier density}.
In particular, a linear resistivity (strange metal) appears over a broad temperature range without saturation (bad metal) with a slope that decreases with increased density.

These exotic behaviors are actually the generic properties of EBL.
Specifically, the temperature, $T$, and density, $x$, dependent low-frequency limit of optical conductivity, $\sigma(\omega\to 0;T,x) \propto \int d^3 k D_{k}(T,x)$, is dominated by
\begin{align}
D_{k}(T,x) &\propto \dot{n}_\mathrm{B}(\epsilon_k) \approx \frac{1}{k_{B}T} \sinh^{-2}\Big(\frac{\epsilon_{k}}{2T}\Big),
\label{D_k}
\end{align}
where $n_\mathrm{B}$ is the Bose-Einstein distribution and $\epsilon_{k}$ the bosonic excitation energy for momentum $k$ measured from the chemical potential $\mu(T,x)$.
Above the vicinity of $T_c$, $\mu(T,x) \approx f(x)-2\lambda T/x$ with a $T$- and $x$-independent coefficient $\lambda$.
In turn, $\frac{\epsilon_{k}}{2T}\to \lambda/x$ becomes $k$-independent, and the resistivity,
\begin{align}
\rho(T>T_c,x)\propto T\sinh^2{(\lambda/x)},
\label{T_linear}
\end{align}
is $T$-linear with a $x$-dependent slope.
On the other hand, for $T$ similar to or lower than $T_c$, $\frac{\epsilon_{k}}{2T}\ll 1$ and $D_k(T\lesssim T_c,x)\propto T/\epsilon_k^2$, such that
\begin{align}
\rho(T\lesssim T_c,x)\propto 1/T,
\label{T_inverse}
\end{align}
instead displays a $T^{-1}$ dependence with weak $x$-dependence.
Altogether, the resistivity is approximately
\begin{align}
\label{Drudeweight}
\rho(T,x) \approx aT^{-1} + b(x) + c(x)T,
\end{align}
with $T$-independent parameters $a$, $b(x)$, and $c(x)$.

For example, the strange metal behavior necessarily results at high enough temperature with Eq.(\ref{T_linear}).
In fact, a very careful analysis of the experimental data confirms that, contrary to popular belief, $T$-linear resistivity is not a low-temperature limit phenomenon.
A clear illustration is the asymptotic $T$-linearity in the observed transverse and longitudinal resistivities in Fig.~\ref{dc_resist}(d), in comparison with our formula Eq.(\ref{Drudeweight}) in (c).
This is in great contrast to many of the existing purely quantum mechanisms that display \textit{low}-temperature power-law scaling (e.g. MFL~\cite{varma}, quantum criticality~\cite{Sachdev1,sachdev_quantum_2011,mousatov_theory_2020}, AdS-CFT~\cite{faulkner_strange_2010,Sachdev_2010}, etc.)

It is crucial to realize that the onset of this ``higher temperature'' physics in the EBL occurs deceivingly at a temperature much \textit{lower} than the full bandwidth.
This does not occur in either Bose gas or one-band systems, where the sum rule strongly constrains the Drude weight and the temperature dependence at this scale.
As elaborated below, this reduced onset temperature scale benefits from two properties of the generic multi-orbital nature of the EBL model: 1) the geometric frustration induced narrowing of the low-energy band, and 2) a more flexible Drude weight allowed by the additional inter-band excitations.

Furthermore, Eq.(\ref{T_linear}) suggests a simple carrier density dependent slope, $\gamma(x)\propto\sinh^2(\frac{\lambda}{x})$, of the $T$-linear resistivity.
Fig.~\ref{dc_resist}(e) compares the observed slopes~\cite{Bozovic1} (red dots) with our numerical result (black squares), both fit remarkably well by Eq.(\ref{T_linear}) (blue curve) up to a \emph{density-independent} constant $\gamma_{0}$ reminiscent of phonon or magnon contributions.

Equation~\ref{T_inverse} also intuitively explains the observed weak insulating behavior, since it becomes more divergent as temperature is further decreased.
In fact, Fig~\ref{dc_resist}(f) shows that a typical observation~\cite{Ando_weakIns_2001} of weak insulating behavior fits Eq.(\ref{Drudeweight}) precisely over multiple decades of temperature range spanning multiple qualitative characteristics. This fitting clearly goes far beyond the accuracy of the popular $\ln(1/T)$ fitting, which covers less than half a decade in the inset of panel (f).
%This a natural result of the normal phase of an EBL, when the energy of quasi-particles exceeds the temperature such that their contributions to the conductivity diminish.

\begin{figure}
	\setlength{\belowcaptionskip}{+0.1cm}
	\begin{center}
		\includegraphics[width=\columnwidth]{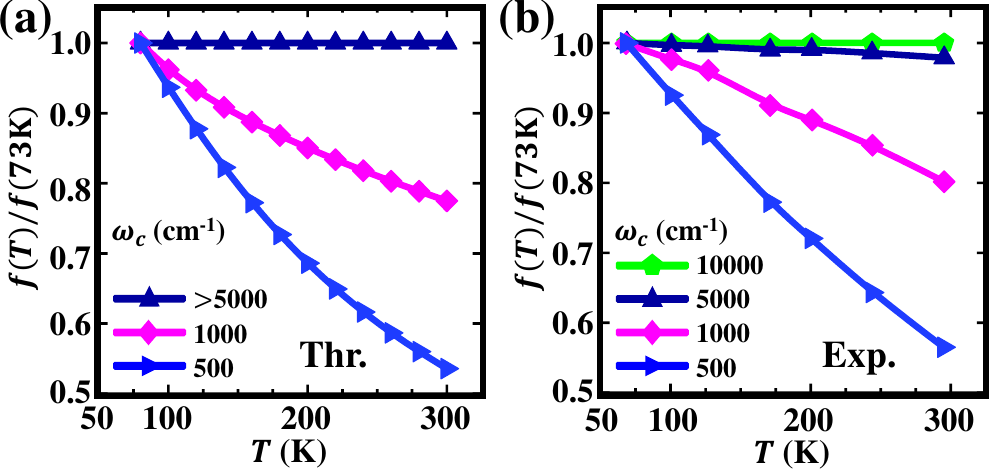}
	\end{center}
	\vspace{-0.5cm}
	\caption{Similar temperature-dependent fraction $f(T)$ of integrated spectral weights in the (a) EBL and (b) observations of Bi-2212~\cite{hwang2007}.
    The spectral weight quickly transfers to higher frequency under increased temperature as indicated by the steep negative slopes for lower cutoff $\omega_{c}$.}
	\label{spec_weight}
	\vspace{-0.5cm}
\end{figure}

Finally, the bad metal behavior with unbounded high-temperature resistivity is inevitable in Eq.(\ref{Drudeweight}).
It turns out that the steadily growing resistivity is a result of the reduction of low-energy carrier density accompanied by a weight transfer to the higher frequency regime.
Indeed, Fig.~\ref{spec_weight}(a) shows (in light blue) that the fraction $f(T)$ of the integrated low-frequency spectral weight, $f=\int_{0}^{\omega_{c}}\sigma d\omega /\int_{0}^{\infty}\sigma d\omega$,  up to some cutoff frequency $\omega_{c}=500$cm$^{-1}$ continuously decreases at higher temperature.
Note that the frequency range of weight transfer is rather large ($>100$meV), since a higher frequency cutoff ($\omega_{c}=1000$cm$^{-1}$) shows a similar trend.
Interestingly, the exact same trend of weight transfer has been observed in recent experiments~\cite{hwang2007} as shown in Fig.~\ref{spec_weight}(b), confirming that the dominant physics for bad metals is the temperature promoted weight transfer.

Not surprisingly, this observation confirms our earlier discussion that a key factor for transport of the EBL is generally its \textit{versatile} low-energy spectral \textit{weight}, a feature much harder to realize under the Pauli principle.
(This is why fermionic transport typically concerns the scattering rate instead.)
In fact, even in Bose gas or one-band bosonic systems~\cite{FWGF,AlexandrovChgBoseFluid,PPRLChg2e,badmetHCB}, without additional spectral features at finite frequency, the Drude weight would be strongly constrained by the optical sum rule.
(Not to mention, without inter-band excitations, the observed mid-infrared peak and feature-less continuum spanning a large energy range would not result.)
It is owing to its multi-orbital nature and bosonic statistics that EBL \textit{generically} displays rich transport features, including the strange metal, bad metal, weak insulator, and a continuum and mid-infrared peak in the spectrum.

Now, since only its essential characteristics matter for our results, none of the features found here are sensitive to parameter fine-tuning or even the underlying fermionic \textit{lattice type}, in direct analogy to the weakly interacting Fermi liquid.
Intuitively, the \textit{simple} particle nature of these `exotic' phenomena within the EBL paradigm offers strong support as a universal low-energy model for a large class of strongly-correlated materials.

The good comparison between the presented EBL in a square lattice and the experimental observations supports its use as an appropriate low-energy model for the charge dynamics in these systems.
Indeed, the known high-energy physics of the cuprates naturally fits into the basic assumptions of the EBL.
The intra-atomic repulsion $\sim 8$eV and the short-range antiferromagnetic correlation $\sim 150$meV are both well-beyond the relevant energy scales in this study and therefore provide adequate high energy constraints.
Furthermore, numerous strong binding mechanisms have been proposed for the these materials with energy scales higher than room temperature, including via two-dimensional short-range antiferromagnetic correlation~\cite{Emery2,Hirsch,Dean} and bi-polaronic correlation~\cite{AlexandrovBipolaron}.
Such agreement suggests that the EBL dominates the charge dynamics of the most actively studied part of the phase diagram.

Finally, the EBL scenario presented here offers great advantage over existing theories of bad metal, strang metal, and weak insulating behavior.
First, our scenario gives what these materials \textit{are} (emergent Bose liquid), instead of focusing on what these materials are \textit{not} (Fermi liquid) via enormous scattering that destroys the particle nature of the carriers~\cite{DMFT1,PhysRevB.91.075124,PhysRevB.94.235115,PhysRevLett.122.186601,faulkner_strange_2010,RevModPhys.68.13,MottQC_2015PRL,SYKpnas, Subir_PRX,Planckian_RevModPhys, cryst2022, Ataei2022,alexandrov_logarithmic_1997,Europhysics2008}.
Second, a generic EBL \textit{simultaneously} hosts the bad metal, strange metal, and sometimes weak insulating behavior, with highly unusual optical features.
This is in great contrast to existing theories~\cite{DMFT1,PhysRevB.91.075124,PhysRevB.94.235115,PhysRevLett.122.186601,faulkner_strange_2010,RevModPhys.68.13,MottQC_2015PRL,SYKpnas, Subir_PRX,Planckian_RevModPhys, cryst2022, Ataei2022,alexandrov_logarithmic_1997,Europhysics2008,Hartnoll2015,varma,Sachdev_2010,Sachdev1,sachdev_quantum_2011,mousatov_theory_2020,Caprara2022,Sachdev2023,YSYK2024,laliberte2016origin}, which account for only one of these characteristics, in spite of their simultaneous occurrence in many materials~\cite{wang_observation_1996,PhysRevB.54.7445,PhysRevLett.69.2975,PhysRevB.42.6342,PhysRevB.41.846,Takenaka,PhysRevLett.124.076801,VSe2_resist2020}.
Third, the EBL naturally host a \textit{correct energy-scale} of its band structure consistent with the experimental observation and comparable to room temperature, owing to the common geometrical frustration of its generic multi-orbital bond lattice~\cite{xzhang}.
This generates rich spectral features correctly in the observed mid-infrared region of the optical conductivity, and naturally allows a strong temperature dependence in nearly all physical properties, particularly the weight transfer essential for the bad metal behavior.
In contrast, the same can only occur at higher energy and temperature in the Hubbard or $t$-$J$ models~\cite{PhysRevB.91.075124,MottQC_2015PRL,RevModPhys.68.13,DMFT1,PhysRevB.91.075124,MottQC_2015PRL,PhysRevB.94.235115,PhysRevLett.122.186601}.
Forth, in good agreement with experimental optical spectra of many materials, temperature dependence of transport for bosonic carriers is mostly through the low-energy spectral weight, instead of the scattering rate as focused in most of existing theories~\cite{DMFT1,PhysRevB.91.075124,PhysRevB.94.235115,PhysRevLett.122.186601,faulkner_strange_2010,RevModPhys.68.13,MottQC_2015PRL,SYKpnas, Subir_PRX,Planckian_RevModPhys, cryst2022, Ataei2022,alexandrov_logarithmic_1997,Europhysics2008}.
Fifth, in EBL the strange metal behavior does not persist to the low-temperature limit, as commonly observed in materials~\cite{Bozovic1,Takenaka,PhysRevLett.124.076801, li_absence_2020, VSe2_resist2020, WenHaihu2018}.
(This, however, invalidates most of the existing theories.)
Finally, the weak insulating behavior is a generic low-temperature behavior of EBL (until the superfluidity takes over), offering an explanation for observations in higher-quality materials.
In contrast, homogeneous fermionic carriers have generally reduced resistivity and thus typically requires various types of disorders~\cite{alexandrov_logarithmic_1997,Europhysics2008} to mimic the weak insulating behavior.

A natural prediction of EBL is the charge 2$e$ carriers for transport in the non-superconducting phase, as detectable via shot noise.
There are currently reports~\cite{2eShotBoz,Bastiaans2eShot} of such observation, but independent confirmation would be highly valuable.
In addition, the bosonic band structure in Fig.~\ref{opt_cond_fig} (b) should be detectable via two-electron coincident photoemission processes~\cite{2e_ARPESsetup,2e_ARPES_TSC}.

In summary, we demonstrate the bad, the strange, and the weak insulating behaviors above the superfluid phase of emergent Bose liquid as a universal paradigm for transport properties in solids complementary to the textbook Fermi liquid theory.
Instead of relying on exotic scattering to destroy the fermionic particle nature, the exploitation of bosonic characteristics combined with a multi-orbital picture allows uncomplicated particle-based mechanisms far from criticality that displays many long-standing puzzles in a wide range of strongly correlated materials.
By providing a very intuitive physical understanding of not only the DC transport behaviors but even the exotic dynamics in the optical conductivity, we help circumvent the need to resort to arcane theoretical description such as that found in `quantum supreme matter'~\cite{ZannenQSM}.
If EBL continues its track record~\cite{weiku1,weiku2,laurence1,hegg1,xander1,xinlei1} as a simple alternative general paradigm to the Fermi liquid, it may eventually rise up as the default description for a wide scope of non-Fermi liquid materials.

\begin{acknowledgments}
We thank Jie Wu, Ruoshi Jiang, Fan Yang, Zhi Wang, Zi-Jian Lang, and Xinlei Yue for useful discussions.
During the pear review process, the preprint of this article were referred to by a review article~\cite{Lang_review} that summarized some of our findings as part of the general properties of EBL.
This work is supported by National Natural Science Foundation of China Grants 12274287 and 12042507, and Innovation Program for Quantum Science and
Technology No. 2021ZD0301900.

%\textcolor{blue}{During the pear review process, the preprint of this article were referred to by a review article~\cite{Lang_review} that summarized some of our findings as part of the general properties of EBL.}
\end{acknowledgments}

\bibliography{optcond}

\end{document}